\documentstyle[a4,11pt]{article} \begin{document}

\begin{titlepage}
\vskip 2cm
\begin{flushright}
Preprint CNLP-1998-01
\end{flushright}
\vskip 2cm
\begin{center}
Soliton equations in N-dimensions as exact reductions of Self-Dual
Yang-Mills equation V. Simplest (2+1)-dimensional soliton equations\footnote{Preprint
CNLP-1998-01. Alma-Ata. 1998.}
\vskip 2cm

{\bf Kur. R. Myrzakul and R. Myrzakulov  }

\end{center}
\vskip 1cm
Centre for Nonlinear Problems, PO Box 30, 480035, Alma-Ata-35, Kazakhstan\\
E-mail: cnlpgnn@satsun.sci.kz

\vskip 1cm

\begin{abstract}
Some aspects of the multidimensional soliton geometry are considered.
It is shown that some simplest (2+1)-dimensional soliton equations are exact
reductions of the  Self-Dual Yang-Mills equation or its higher hierarchy.
\end{abstract}


\end{titlepage}

\setcounter{page}{1}
\tableofcontents

\section{Introduction}
One of the most interesting and important multidimensional integrable equations
is the self-dual Yang-Mills  equation (SDYME)
[2, 27]. This four - dimensional equation arises in relatiivity [26, 3] and in field theory [27].
The SDYM equations describe a connection for a bundle over the
Grassmannian of two-dimensional subspaces of the twistor space.
Integrability for a SDYM connection means that its curvature vanishes
on certain two-planes in the tangent space of the Grassmannian. As
shown in [4,5], This allows one to characterize SDYM connections
in terms of the spliting problem for a transition function in a holomorphic
bundle over the Riemann sphere, i.e. the trivialiization of the bundle
[28, 29].

Resently it has been shown that practically all known soliton equations in
1+1 and 2+1 dimensions may be obtained by reductions of the SDYME [2, 4-5,
18-19, 34-37] (see also the book [1] and references therein).
On the other hand, it is well known that almost all known integrable equations
may be obtained from the some equations of soliton geometry by reductions.
These equations are the integrability conditions of the system describing
the moving orthogonal trihedral of a curve or surface [8-24,38-40,43-44,46].
For example, in 2+1 dimensions, the role of the such geometrical equations
play the mM-LXII or M-LXII equations. In [8] and in our previous notes of
this series [18-21] we have considered some
aspects of the multidimensional soliton geometry (see, also the refs. [38-40,
43-44]).
Also we have studied the relation
between the multidimensional soliton equations and the Self-Dual Yang-Mills
equation. In this note we continue this work.

\section{Soliton geometry in $d=4$ dimensions}

It is well known that exist several equations
describing the 4 - dimensional curves/"surfaces" in n - dimensional space.
Here we present some of them.
\subsection{The M-LXVIII equation}
Consider the M-LXVIII equation [8]
$$
\left( \begin{array}{c}
{\bf e}_{1}\\
{\bf e}_{2}  \\
{\bf e}_{3}  \\
\vdots  \\
{\bf e}_{n}    \\
\end{array} \right)_{\xi_1} = A
\left( \begin{array}{c}
{\bf e}_{1}\\
{\bf e}_{2}  \\
{\bf e}_{3}  \\
\vdots  \\
{\bf e}_{n}    \\
\end{array} \right)_{\xi_3} + B
\left( \begin{array}{c}
{\bf e}_{1}\\
{\bf e}_{2}  \\
{\bf e}_{3}  \\
\vdots  \\
{\bf e}_{n}    \\
\end{array} \right)
\eqno(1a)
$$
$$
\left( \begin{array}{c}
{\bf e}_{1}\\
{\bf e}_{2}  \\
{\bf e}_{3}  \\
\vdots  \\
{\bf e}_{n}    \\
\end{array} \right)_{\xi_2} = C
\left( \begin{array}{c}
{\bf e}_{1}\\
{\bf e}_{2}  \\
{\bf e}_{3}  \\
\vdots  \\
{\bf e}_{n}    \\
\end{array} \right)_{\xi_4} + D
\left( \begin{array}{c}
{\bf e}_{1}\\
{\bf e}_{2}  \\
{\bf e}_{3}  \\
\vdots  \\
{\bf e}_{n}    \\
\end{array} \right)
\eqno(1b)
$$
where ${\bf e}^{2}_{j}=1, ({\bf e}_{i} {\bf e}_{j})=\delta_{ij}$ and
$A(\lambda), B(\lambda), C(\lambda), D(\lambda)$ are (n$\times$n)-matrices,
$\lambda$ is some parameter, $\xi_{i}$ are coordinates. This equation
describes some four dimensional curves and/or "surfaces" in n-dimensional
space. It is one of main equations of the multidimensional soliton geometry
and admits several integrable reductions [8]. The compatibility condition of
these equations gives the M-LXX equation, which contents several interesting
integrable nonlinear evolution equations (NEE).
In this note we will present
some of these integrable reductions.

\subsection{The M-LXXI equation}

Consider the M-LXXI equation [8]
$$
\left( \begin{array}{c}
{\bf e}_{1}\\
{\bf e}_{2}  \\
{\bf e}_{3}  \\
\vdots  \\
{\bf e}_{n}    \\
\end{array} \right)_{\xi_1} = A
\left( \begin{array}{c}
{\bf e}_{1}\\
{\bf e}_{2}  \\
{\bf e}_{3}  \\
\vdots  \\
{\bf e}_{n}    \\
\end{array} \right)
\eqno(2a)
$$
$$
\left( \begin{array}{c}
{\bf e}_{1}\\
{\bf e}_{2}  \\
{\bf e}_{3}  \\
\vdots  \\
{\bf e}_{n}    \\
\end{array} \right)_{\xi_{2}} =  B
\left( \begin{array}{c}
{\bf e}_{1}\\
{\bf e}_{2}  \\
{\bf e}_{3}  \\
\vdots  \\
{\bf e}_{n}    \\
\end{array} \right)
\eqno(2b)
$$
$$
\left( \begin{array}{c}
{\bf e}_{1}\\
{\bf e}_{2}  \\
{\bf e}_{3}  \\
\vdots  \\
{\bf e}_{n}    \\
\end{array} \right)_{\xi_{4}} = C
\left( \begin{array}{c}
{\bf e}_{1}\\
{\bf e}_{2}  \\
{\bf e}_{3}  \\
\vdots  \\
{\bf e}_{n}    \\
\end{array} \right)_{\xi_{3}} + D
\left( \begin{array}{c}
{\bf e}_{1}\\
{\bf e}_{2}  \\
{\bf e}_{3}  \\
\vdots  \\
{\bf e}_{n}    \\
\end{array} \right).
\eqno(2c)
$$

The compatibility condition of these equations gives some NEEs (see, e.g.
the refs. [8,18-19]).

\subsection{The M-LXI equation}
Consider the 4-dimensional M-LXI equation [8]
$$
\left( \begin{array}{c}
{\bf e}_{1}\\
{\bf e}_{2}  \\
{\bf e}_{3}  \\
\vdots  \\
{\bf e}_{n}    \\
\end{array} \right)_{\xi_1} = A
\left( \begin{array}{c}
{\bf e}_{1}\\
{\bf e}_{2}  \\
{\bf e}_{3}  \\
\vdots  \\
{\bf e}_{n}    \\
\end{array} \right)
\eqno(3a)
$$
$$
\left( \begin{array}{c}
{\bf e}_{1}\\
{\bf e}_{2}  \\
{\bf e}_{3}  \\
\vdots  \\
{\bf e}_{n}    \\
\end{array} \right)_{\xi_{2}} =  B
\left( \begin{array}{c}
{\bf e}_{1}\\
{\bf e}_{2}  \\
{\bf e}_{3}  \\
\vdots  \\
{\bf e}_{n}    \\
\end{array} \right)
\eqno(3b)
$$
$$
\left( \begin{array}{c}
{\bf e}_{1}\\
{\bf e}_{2}  \\
{\bf e}_{3}  \\
\vdots  \\
{\bf e}_{n}    \\
\end{array} \right)_{\xi_{3}} = C
\left( \begin{array}{c}
{\bf e}_{1}\\
{\bf e}_{2}  \\
{\bf e}_{3}  \\
\vdots  \\
{\bf e}_{n}    \\
\end{array} \right)
\eqno(3c)
$$
$$
\left( \begin{array}{c}
{\bf e}_{1}\\
{\bf e}_{2}  \\
{\bf e}_{3}  \\
\vdots  \\
{\bf e}_{n}    \\
\end{array} \right)_{\xi_{4}} = D
\left( \begin{array}{c}
{\bf e}_{1}\\
{\bf e}_{2}  \\
{\bf e}_{3}  \\
\vdots  \\
{\bf e}_{n}    \\
\end{array} \right).
\eqno(3d)
$$
The compatibility condition of these equations gives the 4-dimensional
M-LXII equation
$$
A_{\xi_{2}}-B_{\xi_{1}}+[A,B]=0   \eqno(4a)
$$
$$
A_{\xi_{3}}-C_{\xi_{1}}+[A,C]=0   \eqno(4b)
$$
$$
A_{\xi_{4}}-D_{\xi_{1}}+[A,D]=0   \eqno(4c)
$$
$$
C_{\xi_{2}}-B_{\xi_{3}}+[C,B]=0   \eqno(4d)
$$
$$
D_{\xi_{2}}-B_{\xi_{4}}+[D,B]=0   \eqno(4e)
$$
$$
C_{\xi_{4}}-D_{\xi_{3}}+[C,D]=0.   \eqno(4f)
$$
This equation contents many interesting NEEs (see, e.g. the refs. [8,18-19]).

\section{The M-LXX equation}

From (1) we get the following M-LXX equation [8]
$$
AD_{\xi_{3}}-CB_{\xi_{4}}+B_{\xi_{2}}-D_{\xi_{1}}+[B,D]=0
\eqno(5a)
$$
$$
A_{\xi_{2}}-CA_{\xi_{4}}+[A,D]=0
\eqno(5b)
$$
$$
[A,C]=0
\eqno(5c)
$$
$$
C_{\xi_{1}}-AC_{\xi_{3}}+[C,B] =0.
\eqno(5d)
$$

If we choose
$$
A=aI, \quad C=bI, \quad a,b =consts,  \eqno(6)
$$
then the M-LXX equation (5) takes the form
$$
aD_{\xi_{3}}-bB_{\xi_{4}}+B_{\xi_{2}}-D_{\xi_{1}}+[B,D]=0.
\eqno(7)
$$

\section{The SDYME as the particular case of the M-LXX equation}

Now we assume that
$$
B=A_{1}-\lambda A_{3}, \quad D=A_{2}-\lambda A_{4}, \quad a=b=\lambda.
\eqno(8)
$$
So that the M-LXVIII equation (1) takes the form
$$
\left( \begin{array}{c}
{\bf e}_{1}\\
{\bf e}_{2}  \\
{\bf e}_{3}  \\
\vdots  \\
{\bf e}_{n}    \\
\end{array} \right)_{\xi_1} = \lambda
\left( \begin{array}{c}
{\bf e}_{1}\\
{\bf e}_{2}  \\
{\bf e}_{3}  \\
\vdots  \\
{\bf e}_{n}    \\
\end{array} \right)_{\xi_3} + (A_{1}-\lambda A_{3})
\left( \begin{array}{c}
{\bf e}_{1}\\
{\bf e}_{2}  \\
{\bf e}_{3}  \\
\vdots  \\
{\bf e}_{n}    \\
\end{array} \right)
\eqno(9a)
$$
$$
\left( \begin{array}{c}
{\bf e}_{1}\\
{\bf e}_{2}  \\
{\bf e}_{3}  \\
\vdots  \\
{\bf e}_{n}    \\
\end{array} \right)_{\xi_2} = \lambda
\left( \begin{array}{c}
{\bf e}_{1}\\
{\bf e}_{2}  \\
{\bf e}_{3}  \\
\vdots  \\
{\bf e}_{n}    \\
\end{array} \right)_{\xi_4} + (A_{2}-\lambda A_{4})
\left( \begin{array}{c}
{\bf e}_{1}\\
{\bf e}_{2}  \\
{\bf e}_{3}  \\
\vdots  \\
{\bf e}_{n}    \\
\end{array} \right).
\eqno(9b)
$$

From (7) we obtain the SDYME
$$
A_{2\xi_{1}}-A_{1\xi_{2}}+[A_{2},A_{1}]=0  \eqno(10a)
$$
$$
A_{4\xi_{3}}-A_{3\xi_{4}}+[A_{4},A_{3}]=0  \eqno(10b)
$$
$$
A_{1\xi_{4}}-A_{4\xi_{1}}+[A_{1},A_{4}]=
A_{2\xi_{3}}-A_{3\xi_{2}}+[A_{2},A_{3}]  \eqno(10c)
$$
or
$$
F_{\xi_{1}\xi_{2}}=0  \eqno(11a)
$$
$$
F_{\xi_{3}\xi_{4}}=0  \eqno(11b)
$$
$$
F_{\xi_{4}\xi_{1}}-F_{\xi_{3}\xi_{2}}=0.  \eqno(11c)
$$
Here
$$
F_{\xi_{i}\xi_{k}} = A_{k\xi_{i}}-A_{i\xi_{k}}+[A_{k},A_{i}].  \eqno(12)
$$
The SDYME (10) on a connection $A$ are the self-duality conditions
on the curvature under the Hodge star operation
$$
F= \ast F   \eqno(13)
$$
or in index notation
$$
F_{\mu\nu}= \epsilon{\mu\nu\rho\delta}F^{\rho\delta}   \eqno(14)
$$
where $\ast$ is the Hodge operator, $\epsilon_{\mu\nu\rho\delta}$ stands
for the completely antisymmetric tensor in four dimensions with the convention:
$\epsilon_{1234}=1$. The SDYME is integrable by the Inverse Scattering
Transform (IST) method (see, e.g. [2,27]). The Lax representation (LR) of the SDYME has the form [27, 30]
$$
\Phi_{\xi_{1}}-\lambda\Phi_{\xi_{3}}=(A_{1}-\lambda A_{3})\Phi  \eqno(15a)
$$
$$
\Phi_{\xi_{2}}-\lambda\Phi_{\xi_{4}}=(A_{2}-\lambda A_{4})\Phi.  \eqno(15b)
$$

\section{The M-LXVII equation as the particular case of the M-LXVIII
equation}

In this section we consider the 3- dimensional curves/"surfaces".
Let variables in the M-LXVIII equation (1) are independent of $\xi_{3}$. Then
we obtain the following M-LXVII equation [8]
$$
\left( \begin{array}{c}
{\bf e}_{1}\\
{\bf e}_{2}  \\
{\bf e}_{3}  \\
\vdots  \\
{\bf e}_{n}    \\
\end{array} \right)_{\xi_1} = B
\left( \begin{array}{c}
{\bf e}_{1}\\
{\bf e}_{2}  \\
{\bf e}_{3}  \\
\vdots  \\
{\bf e}_{n}    \\
\end{array} \right)
\eqno(16a)
$$
$$
\left( \begin{array}{c}
{\bf e}_{1}\\
{\bf e}_{2}  \\
{\bf e}_{3}  \\
\vdots  \\
{\bf e}_{n}    \\
\end{array} \right)_{\xi_2} = C
\left( \begin{array}{c}
{\bf e}_{1}\\
{\bf e}_{2}  \\
{\bf e}_{3}  \\
\vdots  \\
{\bf e}_{n}    \\
\end{array} \right)_{\xi_4} + D
\left( \begin{array}{c}
{\bf e}_{1}\\
{\bf e}_{2}  \\
{\bf e}_{3}  \\
\vdots  \\
{\bf e}_{n}    \\
\end{array} \right).
\eqno(16b)
$$

In  this case we get the 3-dimensional M-LXX equation
$$
-bB_{\xi_{4}}+B_{\xi_{2}}-D_{\xi_{1}}+[B,D]=0
\eqno(17)
$$
and the 3-dimensional SDYME
$$
A_{2\xi_{1}}-A_{1\xi_{2}}+[A_{2},A_{1}]=0  \eqno(18a)
$$
$$
-A_{3\xi_{4}}+[A_{4},A_{3}]=0  \eqno(18b)
$$
$$
A_{1\xi_{4}}-A_{4\xi_{1}}+[A_{1},A_{4}]=
-A_{3\xi_{2}}+[A_{2},A_{3}].  \eqno(18c)
$$

We note that for the 3-dimensional SDYME (18) the corresponding M-LXVII
equation has the form
$$
\left( \begin{array}{c}
{\bf e}_{1}\\
{\bf e}_{2}  \\
{\bf e}_{3}  \\
\vdots  \\
{\bf e}_{n}    \\
\end{array} \right)_{\xi_1} =
(A_{1}-\lambda A_{3})
\left( \begin{array}{c}
{\bf e}_{1}\\
{\bf e}_{2}  \\
{\bf e}_{3}  \\
\vdots  \\
{\bf e}_{n}    \\
\end{array} \right)
\eqno(19a)
$$
$$
\left( \begin{array}{c}
{\bf e}_{1}\\
{\bf e}_{2}  \\
{\bf e}_{3}  \\
\vdots  \\
{\bf e}_{n}    \\
\end{array} \right)_{\xi_2} = \lambda
\left( \begin{array}{c}
{\bf e}_{1}\\
{\bf e}_{2}  \\
{\bf e}_{3}  \\
\vdots  \\
{\bf e}_{n}    \\
\end{array} \right)_{\xi_4} + (A_{2}-\lambda A_{4})
\left( \begin{array}{c}
{\bf e}_{1}\\
{\bf e}_{2}  \\
{\bf e}_{3}  \\
\vdots  \\
{\bf e}_{n}    \\
\end{array} \right).
\eqno(19b)
$$
So that the curve or "surface" corresponding to the equation (19) is the
integrable.

\section{The Zakharov equation as the exact reduction of the SDYME}

Now let us we consider the gauge condition
$$
A_{4}=0. \eqno(20)
$$
Then the 3-dimensional SDYME takes the form
$$
A_{2\xi_{1}}-A_{1\xi_{2}}+[A_{2},A_{1}]=0  \eqno(21a)
$$
$$
A_{3\xi_{4}}=0  \eqno(21b)
$$
$$
A_{1\xi_{4}}=
-A_{3\xi_{2}}+[A_{2},A_{3}].  \eqno(21c)
$$
If we take
$$
A_{3}=const  \eqno(22)
$$
then the 3-dimensional SDYME has the form
$$
A_{2\xi_{1}}-A_{1\xi_{2}}+[A_{2},A_{1}]=0  \eqno(23a)
$$
$$
A_{1\xi_{4}}=
[A_{2},A_{3}].  \eqno(23b)
$$

Now we consider the case $n=3$. And  we take the following
representations of the connections $A_{1}, A_{2}, A_{3}$
$$
A_{1}=
\left( \begin{array}{ccc}
0  &  i(\phi-r^2 \bar \phi)  &  (\phi+r^2 \bar \phi) \\
-i(\phi-r^2 \bar \phi)  &  0  &  0  \\
-(\phi+r^2 \bar \phi)  &  0  &  0
\end{array} \right)  \eqno(24a)
$$
$$
A_{2}=
\left( \begin{array}{ccc}
0  &  (\phi+r^2 \bar \phi)_y  &  -i(\phi-r^2 \bar \phi)_y \\
-(\phi+r^2 \bar \phi)_y  &  0  &  -v  \\
i(\phi+r^2 \bar \phi)_y  &  v  &  0
\end{array} \right)  \eqno(24b)
$$
$$
A_{3}=
\left( \begin{array}{ccc}
0  &    0  &  0 \\
0  &  0  &  1  \\
0  & -1  &  0
\end{array} \right)  \eqno(24c)
$$
and let
$$
\xi_{1}=x, \quad \xi_{2}=t, \quad \xi_{4}=y.  \eqno(25)
$$
So in our case the equation (23) takes the form
$$
A_{2x}-A_{1t}+[A_{2},A_{1}]=0  \eqno(26a)
$$
$$
A_{1y}=
[A_{2},A_{3}]  \eqno(26b)
$$
or in elements
$$
i\phi_{t}=\phi_{xy}+v\phi  \eqno(27a)
$$
$$
v_{x}=2r^2\partial_{y}|\phi|^2.  \eqno(27b)
$$

It is the Zakharov equation (ZE) [49]. We note that in this case
the corresponding M-LXVII equation (19)  looks like
$$
\left( \begin{array}{c}
{\bf e}_{1}\\
{\bf e}_{2}  \\
{\bf e}_{3}
\end{array} \right)_{x} = (A_{1}-\lambda A_{3})
\left( \begin{array}{c}
{\bf e}_{1}\\
{\bf e}_{2}  \\
{\bf e}_{3}
\end{array} \right)
\eqno(28a)
$$
$$
\left( \begin{array}{c}
{\bf e}_{1}\\
{\bf e}_{2}  \\
{\bf e}_{3}
\end{array} \right)_{t} = \lambda
\left( \begin{array}{c}
{\bf e}_{1}\\
{\bf e}_{2}  \\
{\bf e}_{3}
\end{array} \right)_{y} + A_{2}
\left( \begin{array}{c}
{\bf e}_{1}\\
{\bf e}_{2}  \\
{\bf e}_{3}
\end{array} \right).
\eqno(28b)
$$

From (20), (22) and (28) we get the following LR of the ZE
$$
\Phi_{x}=(A_{1}-\lambda A_{3})\Phi  \eqno(29a)
$$
$$
\Phi_{t}-\lambda\Phi_{y}=A_{2}\Phi  \eqno(29b)
$$
where $A_{i}$ are ${\bf so(3)}$ or ${\bf so(2,1)}$ matrices. It is convenient
to use the well known isomorphism ${\bf so(3)}\simeq {\bf su(2)}$
or ${\bf so(2,1)}\simeq {\bf su(1,1)}$ and to rewrite the equations (29)
in terms of 2$\times$2 matrices.
As result we have the following standard LR of the ZE (27)
$$
\Psi_{x}=U\Psi  \eqno(30a)
$$
$$
\Psi_{t}=\lambda\Psi_{y}+V\Psi  \eqno(30b)
$$
where
$$
U=\frac{i\lambda}{2}\sigma_3 + G, \quad  G=
\left( \begin{array}{cc}
0  &  \phi \\
-r^2 \bar\phi  &  0
\end{array} \right) \eqno(31a)
$$
$$
V=i\sigma_{3} (\frac{1}{2}vI+G_{y}).
\eqno(31b)
$$

We note that in the (1+1)-dimensional case, i.e. when $y=x$ instead of the
equations (28) and (27) we obtain
the (1+1)-dimensional  M-LXVII equation
$$
\left( \begin{array}{c}
{\bf e}_{1}\\
{\bf e}_{2}  \\
{\bf e}_{3}
\end{array} \right)_{x} = (A_{1}-\lambda A_{3})
\left( \begin{array}{c}
{\bf e}_{1}\\
{\bf e}_{2}  \\
{\bf e}_{3}
\end{array} \right)
\eqno(32a)
$$
$$
\left( \begin{array}{c}
{\bf e}_{1}\\
{\bf e}_{2}  \\
{\bf e}_{3}
\end{array} \right)_{t} = [\lambda(A_{1}-\lambda A_{3}) + A_{2}]
\left( \begin{array}{c}
{\bf e}_{1}\\
{\bf e}_{2}  \\
{\bf e}_{3}
\end{array} \right)
\eqno(32b)
$$
and the nonlinear Schrodinger equation
$$
i\phi_{t}=\phi_{xx}+2r^|\phi|^2\phi=0  \eqno(33)
$$

At last we note that if $n>3 $ then we get the n-component generalisation
of the Zakharov equation
$$
i\phi_{jt}=\phi_{jxy}+v\phi_j=0 \eqno(34a)
$$
$$
v_x=2r^2(\sum^{n}_{k=1}|\phi_k|^2)_y \eqno(34b)
$$

\section{Integrable spin systems and
 the SDYME}

Integrable spin systems in 2+1 dimensions also can be considered as exact
reductions of the SDYME. As example we show that Myrzakulov I (M-I)
equation is the reduction
of the SDYME. Consider  the following gauge condition
$$
A_{1}=A_{2}=0. \eqno(35)
$$
In this case  the 3-dimensional SDYME takes the form
$$
-A_{3\xi_{4}}+[A_{4},A_{3}]=0  \eqno(36a)
$$
$$
-A_{4\xi_{1}}=
-A_{3\xi_{2}}  \eqno(36b)
$$
or in  terms of $x,y,t$
$$
-A_{y}+[A_{4},A_{3}]=0  \eqno(37a)
$$
$$
-A_{4x}=
-A_{3t}.  \eqno(37b)
$$
Let  the connections $A_{4},  A_{3}$ have the forms
$$
A_{3}=
\left( \begin{array}{ccc}
0  &  rS_1  & -irS_2 \\
-rS_1  &  0  &  S_3  \\
irS_2  &  -S_3  &  0
\end{array} \right)  \eqno(38a)
$$
$$
A_{4}=
$$
$$
\left( \begin{array}{ccc}
0  & -ir[2iS_3 S_{2y}-2iS_2 S_{3y}+iuS_1] & -r[2S_3 S_{1y}-2S_1 S_{3y}-uS_2] \\
ir[2iS_3 S_{2y}-2iS_2 S_{3y}+iuS_1]  &  0  & -[ir^2(S^{+}S^{-}_{y}-S^{-}S^{+}_y)-uS_3]  \\
r[2S_3 S_{1y}-2S_1 S_{3y}-uS_2   & ir^2(S^{+}S^{-}_{y}-S^{-}S^{+}_{y})-uS_3   &  0
\end{array} \right)  \eqno(38b)
$$
Substituting (38) into (37) we get the M-I equation
$$
iS_{t}= ([S,S_{y}]+2iuS)_{x}   \eqno(39a)
$$
$$
u_{x}=-\frac{1}{2i}tr(S[S_{x},S_{y}]) \eqno(39b)
$$
where
$$
S=
\left( \begin{array}{cc}
S_{3}  &  S^{-} \\
S^{+}  &  -S_{3}
\end{array} \right), \quad S^{\pm}=S_1\pm i S_2.  \eqno(40)
$$

For the M-I equation the corresponding geomerical equation (19)  looks like
$$
\left( \begin{array}{c}
{\bf e}^{\prime}_{1}\\
{\bf e}^{\prime}_{2}  \\
{\bf e}^{\prime}_{3}
\end{array} \right)_{x} = -\lambda A_{3}
\left( \begin{array}{c}
{\bf e}^{\prime}_{1}\\
{\bf e}^{\prime}_{2}  \\
{\bf e}^{\prime}_{3}
\end{array} \right)
\eqno(41a)
$$
$$
\left( \begin{array}{c}
{\bf e}^{\prime}_{1}\\
{\bf e}^{\prime}_{2}  \\
{\bf e}^{\prime}_{3}
\end{array} \right)_{t} = \lambda
\left( \begin{array}{c}
{\bf e}^{\prime}_{1}\\
{\bf e}^{\prime}_{2}  \\
{\bf e}^{\prime}_{3}
\end{array} \right)_{y} -\lambda  A_{4}
\left( \begin{array}{c}
{\bf e}^{\prime}_{1}\\
{\bf e}^{\prime}_{2}  \\
{\bf e}^{\prime}_{3}
\end{array} \right)
\eqno(41b)
$$
and called the M-LXVI equation. As known [8], the M-LXVII and M-LXVI equations
(41), (28) and (19) are some integrable
(2+1)-dimensional extensions of the Serret-Frenet equation (SFE)
$$
\left( \begin{array}{c}
{\bf e}^{\prime}_{1}\\
{\bf e}^{\prime}_{2}  \\
{\bf e}^{\prime}_{3}
\end{array} \right)_{x} =
\left( \begin{array}{ccc}
0  &  k  &  0\\
-\beta k &  0  &  \tau   \\
0  & -\tau&  0
\end{array} \right)
\left( \begin{array}{c}
{\bf e}^{\prime}_{1}\\
{\bf e}^{\prime}_{2}  \\
{\bf e}^{\prime}_{3}
\end{array} \right) \eqno(42)
$$
or the Codazzi-Mainardi equation (CME) for the surfaces.

The LR of the M-I equation follows from the LR of the SDYME and has the form
$$
\Phi^{\prime}_{x}=-\lambda A_{3}\Phi^{\prime}  \eqno(43a)
$$
$$
\Phi^{\prime}_{t}-\lambda\Phi^{\prime}_{y}=-\lambda A_{4}\Phi^{\prime} .
 \eqno(43b)
$$
As for ZE, we can rewrite the LR of the M-I equation in the standart form,
in terms of 2$\times$2 -  matrices
$$
\Psi^{\prime}_{x}=U^{\prime}\Psi^{\prime}  \eqno(44a)
$$
$$
\Psi^{\prime}_{t}=\lambda\Psi^{\prime}_{y}+V^{\prime}\Psi^{\prime}  \eqno(44b)
$$
where
$$
U^{\prime}=\frac{i\lambda}{2}S
\eqno(45a)
$$
$$
V^{\prime}=\frac{\lambda}{4} ([S,S_{y}]+2iuS).
\eqno(45b)
$$

As above for the ZE (27), in the 1+1 dimensions $(y=x)$ instead of the equations
(41) and (39), we get the
(1+1)-dimensional M-LXVI  equation
$$
\left( \begin{array}{c}
{\bf e}^{\prime}_{1}\\
{\bf e}^{\prime}_{2}  \\
{\bf e}^{\prime}_{3}
\end{array} \right)_{x} = -\lambda A_{3}
\left( \begin{array}{c}
{\bf e}^{\prime}_{1}\\
{\bf e}^{\prime}_{2}  \\
{\bf e}^{\prime}_{3}
\end{array} \right)
\eqno(46a)
$$
$$
\left( \begin{array}{c}
{\bf e}^{\prime}_{1}\\
{\bf e}^{\prime}_{2}  \\
{\bf e}^{\prime}_{3}
\end{array} \right)_{t} =-(\lambda^2 A_3
+\lambda  A_{4})
\left( \begin{array}{c}
{\bf e}^{\prime}_{1}\\
{\bf e}^{\prime}_{2}  \\
{\bf e}^{\prime}_{3}
\end{array} \right)
\eqno(46b)
$$
and the Landau-Lifshitz equation
$$
iS_{t}= \frac{1}{2}[S,S_{xx}].   \eqno(47a)
$$

\section{Gauge equivalence}

It is well known that the ZE (33) and the M-I equation (39) are gauge and
Lakshmanan equivalent
(G-equivalent and L-equivalent) to each other. In our case this fact realize by
the transformation
$$
\left( \begin{array}{c}
{\bf e}^{\prime}_{1}\\
{\bf e}^{\prime}_{2}  \\
{\bf e}^{\prime}_{3}
\end{array} \right) = G
\left( \begin{array}{c}
{\bf e}_{1}\\
{\bf e}_{2}  \\
{\bf e}_{3}
\end{array} \right)
\eqno(48)
$$
or
$$
\Phi^{\prime}=h^{-1}\Phi  \eqno(49a)
$$
or
$$
\Psi^{\prime}=h^{-1}\Psi  \eqno(50)
$$
where $h$ is the solution of the equations (29) or (30) as $\lambda=0$.

\section{A nonisospectral case and the breaking solutions of the SDYME}

Usually for the SDYME (10) the spectral parameter $\lambda$=constant. But
in general it satisfies the following set of nonlinear equations
$$
\lambda_{\xi_{1}}=\lambda\lambda_{\xi_{3}} \eqno(51a)
$$
$$
\lambda_{\xi_{2}}=\lambda\lambda_{\xi_{4}}. \eqno(51b)
$$
These equations have the following solutions
$$
\lambda=\frac{n_{1}\xi_{3}+n_{3}}{n_{4}-n_{1}\xi_{1}}  \eqno(52a)
$$
$$
\lambda=\frac{m_{1}\xi_{4}+m_{3}}{m_{4}-m_{1}\xi_{2}}.  \eqno(52b)
$$
So that the general solution of the set (51) has the form
$$
\lambda=\frac{n_{1}\xi_{3}+n_{3}+m_{1}\xi_{4}}{n_{4}-n_{1}\xi_{1}-
m_{1}\xi_{2}}  \eqno(53)
$$
where $m_{i}, n_{i} =constants$. The corresponding solution of the SDYME (10)
is called the breaking (overlapping) solutions. In the case (18), i.e. for
the ZE
and the M-I equation the set of equations (21) takes the form
$$
\lambda_{x}=0 \eqno(54a)
$$
$$
\lambda_{t}=\lambda\lambda_{y} \eqno(54b)
$$
and the solution (9) has the form
$$
\lambda=\frac{n_{1}y +n_{3}}{n_{4}-n_{1}t}.  \eqno(55)
$$

\section{Hierarchy of the M-LXVIII and Self-Dual Yang-Mills  equations}
The higher hierarchy of the M-LXVIII equation for the SDYME case we write
in the form
$$
\left( \begin{array}{c}
{\bf e}_{1}\\
{\bf e}_{2}  \\
{\bf e}_{3}  \\
\vdots  \\
{\bf e}_{n}    \\
\end{array} \right)_{\xi_1} =
\sum^{k_{1}}_{i=0}A_{i}\lambda^{i}
\left( \begin{array}{c}
{\bf e}_{1}\\
{\bf e}_{2}  \\
{\bf e}_{3}  \\
\vdots  \\
{\bf e}_{n}    \\
\end{array} \right)_{\xi_3} +
\sum^{k_{2}}_{i=0}B_{i}\lambda^{i}
\left( \begin{array}{c}
{\bf e}_{1}\\
{\bf e}_{2}  \\
{\bf e}_{3}  \\
\vdots  \\
{\bf e}_{n}    \\
\end{array} \right)
\eqno(56a)
$$
$$
\left( \begin{array}{c}
{\bf e}_{1}\\
{\bf e}_{2}  \\
{\bf e}_{3}  \\
\vdots  \\
{\bf e}_{n}    \\
\end{array} \right)_{\xi_2} =
\sum^{k_{3}}_{i=0}C_{i}\lambda^{i}
\left( \begin{array}{c}
{\bf e}_{1}\\
{\bf e}_{2}  \\
{\bf e}_{3}  \\
\vdots  \\
{\bf e}_{n}    \\
\end{array} \right)_{\xi_4} +
\sum^{k_{4}}_{i=0}D_{i}\lambda^{i}
\left( \begin{array}{c}
{\bf e}_{1}\\
{\bf e}_{2}  \\
{\bf e}_{3}  \\
\vdots  \\
{\bf e}_{n}    \\
\end{array} \right).
\eqno(56b)
$$

The compatibility condition of these equations yields
the higher hierarchy SDYME. As example, we consider the 3-dimensional case,
work the notation (25) and $n=3$.  Then insteod of (56) we obtain the M-LXVII
equation in the form
$$
\left( \begin{array}{c}
{\bf e}_{1}\\
{\bf e}_{2}  \\
{\bf e}_{3}
\end{array} \right)_{x} =
\sum^{k_{2}}_{i=0}B_{i}\lambda^{i}
\left( \begin{array}{c}
{\bf e}_{1}\\
{\bf e}_{2}  \\
{\bf e}_{3}
\end{array} \right)
\eqno(57a)
$$
$$
\left( \begin{array}{c}
{\bf e}_{1}\\
{\bf e}_{2}  \\
{\bf e}_{3}
\end{array} \right)_{t} =
\sum^{k_{3}}_{i=0}C_{i}\lambda^{i}
\left( \begin{array}{c}
{\bf e}_{1}\\
{\bf e}_{2}  \\
{\bf e}_{3}
\end{array} \right)_{y} +
\sum^{k_{4}}_{i=0}D_{i}\lambda^{i}
\left( \begin{array}{c}
{\bf e}_{1}\\
{\bf e}_{2}  \\
{\bf e}_{3}
\end{array} \right).
\eqno(57b)
$$

It is remarkable that using the equation (57) we can show that some known
(2+1)-dimensional soliton equations are exact reductions of the higher
hierarchy of the SDYME. Here we present some of them.

\subsection{The (2+1)-dimensional mKdV equation}.

In the equation (57)  we assume that
$$
k_{2}=1, k_{3}=2, k_{4}=2, C_{2}=1,
\quad C_{1}=C_{0}=0. \eqno(58)
$$
Thus in this case the M-LXVII equation looks like
$$
\left( \begin{array}{c}
{\bf e}_{1}\\
{\bf e}_{2}  \\
{\bf e}_{3}
\end{array} \right)_{x} = (A_{1}-\lambda A_{3})
\left( \begin{array}{c}
{\bf e}_{1}\\
{\bf e}_{2}  \\
{\bf e}_{3}
\end{array} \right)
\eqno(59a)
$$
$$
\left( \begin{array}{c}
{\bf e}_{1}\\
{\bf e}_{2}  \\
{\bf e}_{3}
\end{array} \right)_{t} = \lambda^{2}
\left( \begin{array}{c}
{\bf e}_{1}\\
{\bf e}_{2}  \\
{\bf e}_{3}
\end{array} \right)_{y} + (D_{2}\lambda^{2}+D_{1}\lambda +D_{0})
\left( \begin{array}{c}
{\bf e}_{1}\\
{\bf e}_{2}  \\
{\bf e}_{3}
\end{array} \right)
\eqno(59b)
$$
where
$$
A_{1}=
\left( \begin{array}{ccc}
0  &  -(q+p)  &  i(q-p) \\
q+p &  0  &  0  \\
-i(q-p) & 0 & 0
\end{array} \right)
\eqno (60)
$$
and $A_3, D_k $ are some matrices [8].
Then the complex functions $q,p$
satisfy the (2+1)-dimensional complex mKdV equation
$$
q_{t} +q_{xxy}-(qv_{1})_{x}-qv_{2}=0,    \eqno(61a) $$
$$ v_{1x}=2E(\bar q q)_{y}    \eqno(61b) $$
$$ v_{2x}=2E(\bar q q_{xy}-\bar q_{xy}q)    \eqno(61c) $$
If $p=q$ is real,
we get the following (2+1)- dimensional mKdV equation
$$
q_{t} +q_{xxy}-(qv_{1})_{x}=0,    \eqno(62a)
$$
$$
v_{1x}=2E(q^2)_{y}=4Eqq_y.    \eqno(62b)
$$

\subsection{The (2+1)-dimensional derivative NLSE}.

In  (57) now   we put
$$
k_{2}=k_{3}=k_{4}=2; C_{2}=2c, C_1=C_0=0, \quad B_{2}=-cA_{3},
 \quad B_{2}=2cA_{1}.
 \eqno(63)
$$
So that the M-LXVII equation takes the form
$$
\left( \begin{array}{c}
{\bf e}_{1}\\
{\bf e}_{2}  \\
{\bf e}_{3}
\end{array} \right)_{x} = (A_{3}\lambda^{2}+ A_{1}\lambda)
\left( \begin{array}{c}
{\bf e}_{1}\\
{\bf e}_{2}  \\
{\bf e}_{3}
\end{array} \right)
\eqno(64a)
$$
$$
\left( \begin{array}{c}
{\bf e}_{1}\\
{\bf e}_{2}  \\
{\bf e}_{3}
\end{array} \right)_{t} = \lambda^{2}
\left( \begin{array}{c}
{\bf e}_{1}\\
{\bf e}_{2}  \\
{\bf e}_{3}
\end{array} \right)_{y} + (D_{2}\lambda^{2}+D_{1}\lambda +D_{0})
\left( \begin{array}{c}
{\bf e}_{1}\\
{\bf e}_{2}  \\
{\bf e}_{3}
\end{array} \right)
\eqno(64b)
$$
where $A_{1}$ is given by (60).
Then the complex functions $q,p$
satisfy the (2+1)-dimensional derivative NLSE
$$
iq_{t}=q_{xy}-2ic(vq)_x  \eqno(65a)
$$
$$
-ip_{t}=p_{xy}+2ic(vq)_x  \eqno(65b)
$$
$$
v_{x}=2(pq)_{y}  \eqno(65c)
$$
which is the Strachan equation [45].

\subsection{The M-III$_{q}$ equation}.

Now we consider the case
$$
k_{2}=k_{3}=k_{4}=2, C_{2}=2cI,C_1=2dI,C_0=0, B_{2}=-cA_{3},
 B_{1}=-dA_3+2cA_1, B_0=dA_{1}
 \eqno(66)
$$
for which the M-LXVII equation has the form
$$
\left( \begin{array}{c}
{\bf e}_{1}\\
{\bf e}_{2}  \\
{\bf e}_{3}
\end{array} \right)_{x} = [{A_{3}(c\lambda^{2}+d\lambda)+ A_{1}(2c\lambda+d)}]
\left( \begin{array}{c}
{\bf e}_{1}\\
{\bf e}_{2}  \\
{\bf e}_{3}
\end{array} \right)
\eqno(67a)
$$
$$
\left( \begin{array}{c}
{\bf e}_{1}\\
{\bf e}_{2}  \\
{\bf e}_{3}
\end{array} \right)_{t} = 2(c\lambda^{2}+d\lambda)
\left( \begin{array}{c}
{\bf e}_{1}\\
{\bf e}_{2}  \\
{\bf e}_{3}
\end{array} \right)_{y} + (D_{2}\lambda^{2}+D_{1}\lambda +D_{0})
\left( \begin{array}{c}
{\bf e}_{1}\\
{\bf e}_{2}  \\
{\bf e}_{3}
\end{array} \right)
\eqno(67b)
$$
where $A_{1}$ is given by (60).
Then the complex functions $q,p$
satisfy the (2+1)-dimensional M-III$_{q}$ equation [8]
$$
iq_{t}=q_{xy}-2ic(vq)_{x}+d^{2}vq  \eqno(68a)
$$
$$
-ip_{t}=p_{xy}+2ic(vq)_{x}+d^{2}vp  \eqno(68b)
$$
$$
v_{x}=2(pq)_{y}.  \eqno(68c)
$$
The M-III$_{q}$ equation (68) admits two integrable reductions: the
Strachan equation (65) as $d=0$ and the ZE (27) as $c=0$.
If we rewrite the equations (67) in terms of $2\times 2$ - matrices, we get
the following LR of the M-III$_{q}$ equation
$$
\Psi_{x}=U\Psi  \eqno(69a)
$$
$$
\Psi_{t}=2(c\lambda^{2}+d\lambda)\Psi_{y}+V\Psi  \eqno(69b)
$$
where
$$
U=i[(c\lambda^{2}+d\lambda)\sigma_{3} + (2c\lambda+d)G], \quad  G=
\left( \begin{array}{cc}
0  &  q \\
p  &  0
\end{array} \right) \eqno(70a)
$$
$$
V=B_{2}\lambda^{2}+B_{1}\lambda+B_{0}.
\eqno(70b)
$$
Here
$$
B_{0}=\frac{d}{2c}B_{1}-\frac{d^{2}}{4c^{2}}B_{2}, \quad
B_{2}=--2ic^{2}v\sigma_{3},
$$
$$
B_{1}=-2icdv\sigma_{3}+2cG_{y}\sigma_{3}-4ic^{2}vG. \eqno(71)
$$

\subsection{The M-XXII$_{q}$ equation}
Now let
$$
k_2=k_3=k_4=2. B_2=A_3, B_1=A_1, B_0=\frac{pq}{4}A_3,
C_2=2I, C_1=C_0=0. \eqno(72)
$$
Then the functions $q,p$ satisfy the following $M-XXII_q$ equations [8]
$$
iq_t + q_{yx} + \frac{i}{2}[(v_1q)_x - v_2 q - qpq_y] = 0 \eqno(73a)
$$
$$
ip_t - p_{yx} + \frac{i}{2}[(v_1p)_x + v_{2}p - qpp_y] = 0 \eqno(73b) $$
$$
v_{1x}=(pq)_y \eqno(73c)
$$
$$
v_{2x}=p_{yx}q-pq_{yx}. \eqno(73d)
$$
This  set of equations is the   G- and
L-equivalent counterpart
of the M-XXII$_{s}$ equation (spin system)[8,25]. The LR
of this equation has the form
$$
\Psi_{2x} = \{ - i(\lambda^{2} - \frac{pq}{4})\sigma_{3} +
\lambda Q\}\Psi_{2} \eqno(74a)
$$
$$
\Psi_{2t} = 2\lambda^{2}\Psi_{2y} + (\lambda^{2} B_{2}  +  \lambda B_{1}
+ B_{0})\Psi_{2}  \eqno(74b)
$$
with
$$
Q =
\left ( \begin{array}{cc}
0   & q \\
-p  & 0
\end{array} \right) ,
B_{2} = \frac{i}{2}v_{1}\sigma_{3},\quad
B_{1} =  i\sigma_{3}Q_{y} - \frac{1}{2}v_{1}Q,\quad
B_{0} = \frac{1}{4}v_{2} - \frac{i}{8}pqv_{1}. \eqno(75)
$$

Now let us consider the following transformation
$$
q^{\prime} = q\exp[-\frac{i}{2}\partial^{-1}_{x}(pq)],
p^{\prime}=p\exp[\frac{i}{2} \partial^{-1}_{x}(pq)].   \eqno(76)
$$

Then the new variables $p^{\prime},q^{\prime}$ satisfy the Strachan
equation [45]
$$ iq^{\prime}_{t} + q^{\prime}_{xy} + i(v^{\prime}q^{\prime})_x = 0, \eqno(77a)$$
$$ ip^{\prime}-p^{\prime}_{xy}+i(v^{\prime}p^{\prime})=0, \eqno(77b) $$
$$ v^{\prime}_x = E(p^{\prime}q^{\prime})_y. \eqno (77c) $$

We see that  the M-XXII$_{q}$ equation (73) and the Strachan equation (70) is
gauge eqivalent to each other. The tranformation (76) induces the
following tranformation of the Jost function$\Psi_{1}$
$$ \Psi_{1} = f^{-1} \Psi_{2} \eqno(78) $$
where $\Psi_{1}$ is the solution of the equations (30) as $d=0$ and
$$
f = \exp(-\frac{i}{4}\partial^{-1}_{x}\mid q\mid^{2}\sigma_{3}) =
\Psi^{-1}_{1}\mid_{\lambda=0}.   \eqno(79)
$$
Then the new Jost function $\Psi_{2} $ satisfies the
following set of equations
$$
\Psi_{2x} = \{ - i\lambda^{2} \sigma_{3} +
\lambda Q^{\prime} \}\Psi_{2}
\eqno(80a)
$$
$$
\Psi_{2t} = 2\lambda^{2}\Psi_{2y} + \{\lambda^{2} B^{\prime}_{2}  +
\lambda B^{\prime}_{1} + B^{\prime}_{0})\}\Psi_{2}  \eqno(80b)
$$
with
$$
Q =
\left ( \begin{array}{cc}
0            & q^{\prime} \\
-p^{\prime}  & 0
\end{array} \right)  \eqno(81)
$$
and $B^{\prime}_{j}$ are given in [8,25,46]. Note that the Strachan (77),
(65) and
MXXII$_{q}$ (73) equations are the simplest (2+1)-dimensional extensions of
the following known NLSE
$$
iq_t + q_{xx} + i(pq^{2})_x  = 0 \eqno(82a)
$$
$$
ip_t - p_{xx} + i(qp^{2})_{x}=0 \eqno(82b) $$
and
$$
iq_{t} + q_{xx} + ipqq_x = 0 \eqno(83a)
$$
$$
ip_{t}-p_{xx}+ipqp_{x}=0 \eqno(83b)
$$
respectively. It is well known that these equations are gauge equivalent
to each other [47].

\section{The M-LXI and M-LXII equations and soliton equations
in $d=3$ dimensions}

\subsection{The M-LXI equation}
The M-LXI equation in $d=3$ dimensions has the form
$$
\left( \begin{array}{c}
{\bf e}_{1}\\
{\bf e}_{2}  \\
{\bf e}_{3}  \\
\vdots  \\
{\bf e}_{n}    \\
\end{array} \right)_{\xi_1} = A
\left( \begin{array}{c}
{\bf e}_{1}\\
{\bf e}_{2}  \\
{\bf e}_{3}  \\
\vdots  \\
{\bf e}_{n}    \\
\end{array} \right)
\eqno(84a)
$$
$$
\left( \begin{array}{c}
{\bf e}_{1}\\
{\bf e}_{2}  \\
{\bf e}_{3}  \\
\vdots  \\
{\bf e}_{n}    \\
\end{array} \right)_{\xi_{2}} =  B
\left( \begin{array}{c}
{\bf e}_{1}\\
{\bf e}_{2}  \\
{\bf e}_{3}  \\
\vdots  \\
{\bf e}_{n}    \\
\end{array} \right)
\eqno(84b)
$$
$$
\left( \begin{array}{c}
{\bf e}_{1}\\
{\bf e}_{2}  \\
{\bf e}_{3}  \\
\vdots  \\
{\bf e}_{n}    \\
\end{array} \right)_{\xi_{3}} = C
\left( \begin{array}{c}
{\bf e}_{1}\\
{\bf e}_{2}  \\
{\bf e}_{3}  \\
\vdots  \\
{\bf e}_{n}    \\
\end{array} \right).
\eqno(84c)
$$
\subsection{The M-LXII equation}
The compatibility condition of the equations (84) gives the 3-dimensional
M-LXII equation
$$
A_{\xi_{2}}-B_{\xi_{1}}+[A,B]=0   \eqno(85a)
$$
$$
A_{\xi_{3}}-C_{\xi_{1}}+[A,C]=0   \eqno(85b)
$$
$$
C_{\xi_{2}}-B_{\xi_{3}}+[C,B]=0.   \eqno(85c)
$$

This equation is the particular case of the Bogomolny equation (BE) [1]
$$
\Psi_{\xi_{3}}+[\Psi,C]+A_{\xi_{2}}-B_{\xi_{1}}+[A,B]=0   \eqno(86a)
$$
$$
\Psi_{\xi_{2}}+[\Psi,B]+A_{\xi_{3}}-C_{\xi_{1}}+[A,C]=0   \eqno(86b)
$$
$$
\Psi_{\xi_{1}}+[\Psi,A]+C_{\xi_{2}}-B_{\xi_{3}}+[C,B]=0.   \eqno(86c)
$$
In fact, frome hence as $\Psi=0$ we obtain the M-LXII equation (85). As well
known that the BE (86) is integrable (see, e.g. the book [1]). As the
particular case of the integrable BE (86), the M-LXII equation is also
integrable.
The corresponding LR has the form
$$
\Phi_{\xi_{1}}-\lambda\Phi_{\xi_{3}}=[-iC -\lambda(A+iB)]\Phi  \eqno(87a)
$$
$$
\Phi_{\xi_{2}}-\lambda\Phi_{\xi_{4}}=[A-iB -\lambda iC]\Phi.  \eqno(87b)
$$
Let us consider the case $n=3, \xi_{1}=x, \xi_{2}=y, \xi_{3}=t$. Then  the
mM-LXI amd mM-LXII equations take the form
$$
\left( \begin{array}{c}
{\bf e}_{1}\\
{\bf e}_{2}  \\
{\bf e}_{3}
\end{array} \right)_{x} = A
\left( \begin{array}{c}
{\bf e}_{1}\\
{\bf e}_{2}  \\
{\bf e}_{3}
\end{array} \right)
\eqno(88a)
$$
$$
\left( \begin{array}{c}
{\bf e}_{1}\\
{\bf e}_{2}  \\
{\bf e}_{3}
\end{array} \right)_{y} =  B
\left( \begin{array}{c}
{\bf e}_{1}\\
{\bf e}_{2}  \\
{\bf e}_{3}
\end{array} \right)
\eqno(88b)
$$
$$
\left( \begin{array}{c}
{\bf e}_{1}\\
{\bf e}_{2}  \\
{\bf e}_{3}
\end{array} \right)_{t} = C
\left( \begin{array}{c}
{\bf e}_{1}\\
{\bf e}_{2}  \\
{\bf e}_{3}
\end{array} \right)
\eqno(88c)
$$
and
$$
A_{y}-B_{x}+[A,B]=0   \eqno(89a)
$$
$$
A_{t}-C_{x}+[A,C]=0   \eqno(89b)
$$
$$
C_{y}-B_{t}+[C,B]=0   \eqno(89c)
$$
where
$$
A =
\left ( \begin{array}{ccc}
0             & k     &-\sigma \\
-\beta k      & 0     & \tau  \\
\beta\sigma   & -\tau & 0
\end{array} \right) ,
\quad
B=
\left ( \begin{array}{ccc}
0            & m_{3}  & -m_{2} \\
-\beta m_{3} & 0      & m_{1} \\
\beta m_{2}  & -m_{1} & 0
\end{array} \right)
$$
$$
D=
\left ( \begin{array}{ccc}
0            & \omega_{3}  & -\omega_{2} \\
-\beta \omega_{3} & 0      & \omega_{1} \\
\beta \omega_{2}  & -\omega_{1} & 0
\end{array} \right).  \eqno(90)
$$

The mM-LXII  equation contents many (and perhaps all?)
known soliton equations (see, e.g. the refs.[8,18-24,38-40,43-44]). We note
that in the case $\sigma=0$ (89) is called the M-LXII equation. Some examples
as follows.

\subsubsection{The Ishimori and DS equations}

In this section, we obtain the Ishimori (IE) and DS equations from
the M-LXI and M-LXII $(\sigma=0)$ equations as some exact reductions.
The IE reads as
$$
{\bf S}_{t}  =  {\bf S}\wedge ({\bf S}_{xx} +\alpha^2 {\bf S}_{yy})+
u_x{\bf S}_{y}+_y{\bf S}_{x} \eqno (91a)
$$
$$
u_{xx}-\alpha^2 u_{yy}  = -2\alpha^2 {\bf S}\cdot ({\bf S}_{x}\wedge
                        {\bf S}_{y}).   \eqno (91b)
$$
We take the following identification
$$
{\bf S}={\bf e}_{1}. \eqno(92)
$$
In this case we have
$$
m_{1}=\partial_{x}^{-1}[\tau_{y}-\frac{\beta}{2\alpha^2}M_2^{Ish}u] \eqno(93a)
$$
$$
m_{2}= -\frac{1}{2\alpha^2 k}M_2^{Ish}u \eqno (93b)
$$
$$
m_{3} =\partial_{x}^{-1}[k_y +\frac{\tau}{2\alpha^2 k}M_2^{Ish}u] \eqno(93c)
$$
$$
M^{IE}_{2}u=u_{xx}-\alpha^{2}u_{yy} \eqno(94)
$$
and
$$
\omega_{1} = \frac{1}{k}[-\omega_{2x}+\tau\omega_{3}]
                 \eqno (95a)
$$
$$
\omega_{2}= -k_{x}-
 \alpha^{2}(m_{3y}+m_{2}m_{1})+im_{2}u_{x}
\eqno (95b)
$$
$$
\omega_{3}= -k \tau+\alpha^{2}(m_{2y}-m_{3}m_{1})
+ik u_{y}+im_{3}u_{x}.
\eqno (95c)
$$
Now let us introduce two complex functions $q, p$, according to the formulae
$$
q = a_{1}e^{ib_{1}}, \quad p=a_{2}e^{ib_{2}}. \eqno(96)
$$
Let $a_{j}, b_{j}$ have the forms
$$
a_{1}^2 =\frac{1}{4}k^2+
\frac{|\alpha|^2}{4}(m_3^2 +m_2^2)-\frac{1}{2}\alpha_{R}km_3-
\frac{1}{2}\alpha_{I}km_2
  \eqno(97a)
$$
$$
b_1 =\partial_{x}^{-1}\{-\frac{\gamma_1}{2ia_{1}^{2}}-
(\bar A-A+D-\bar D)\}  \eqno(97b)
$$
$$
a_2^2=\frac{1}{4}k^2+
\frac{|\alpha|^2}{4}(m_3^2 +m_2^2)+\frac{1}{2}\alpha_{R}km_3
-\frac{1}{2}\alpha_{I}km_2
  \eqno(97c)
$$
$$
b_{2} =\partial_{x}^{-1}\{-\frac{\gamma_2}{2ia_2^{2}}-
(A-\bar A+\bar D-D)\}  \eqno(97d)
$$
where
$$
\gamma_1=i\{\frac{1}{2}k^{2}\tau+
\frac{|\alpha|^2}{2}(m_3km_1+m_2k_y)-
$$
$$
\frac{1}{2}\alpha_{R}(k^{2}m_1+m_3k\tau+
m_2k_x)
+\frac{1}{2}\alpha_{I}[k(2k_y-m_{3x})-
k_x m_3]\}. \eqno(98a)
$$
$$
\gamma_2=-i\{\frac{1}{2}k^{2}\tau+
\frac{|\alpha|^2}{2}(m_3km_1+m_2 k_y)+
$$
$$
\frac{1}{2}\alpha_{R}(k^{2}m_1+m_3k\tau+
m_2k_x )
+\frac{1}{2}\alpha_{I}[k(2k_y-m_{3x})-
k_x m_3]\}. \eqno(98b)
$$
Here $\alpha=\alpha_{R}+i\alpha_{I}$. In this case, $q,p$ satisfy the following
DS  equation
$$
iq_t  + q_{xx}+\alpha^{2}q_{yy} + vq = 0 \eqno (99a)
$$
$$
-ip_t +  p_{xx}+\alpha^{2}p_{yy} + vp = 0 \eqno (99b)
$$
$$
v_{xx}-\alpha^{2}v_{yy} + 2[(p q)_{xx}+\alpha^{2}(p q)_{yy}] = 0.
\eqno (99c)
$$
It is means that  the IE (91) and the DS equation (99) are
L-equivalent [44] to each other.
As well known that these equations are G-equivalent to each other [48].
A few comments are in order.

i) From these results, we get the Ishimori I and DS-I equations as
$\alpha_{R}=1, \alpha_{I}=0$

ii) the Ishimori II and DS-II equations as $\alpha_{R}=0, \alpha_{I}=1$.

iii) For  DS-II equation we have
$$
pq=\mid q\mid^{2}=\mid p\mid^{2}  \eqno(100)
$$

iv) at the same time, for the DS-I equation we obtain
$$
pq\neq\mid q\mid^{2}\neq\mid p\mid^{2}  \eqno(101)
$$
$$
\mid q\mid^{2}=\mid p\mid^{2}-km_{3}  \eqno(102)
$$
$$
pq= (pq)_{R}+i(pq)_{I}
\eqno(103)
$$
so that  $pq$ is the complex quantity.

\subsubsection{The KP and M-X  equations}

The (2+1)-dimensional mM-LXI equation in plane has the form [8]
$$
\left ( \begin{array}{cc}
{\bf e}_{1} \\
{\bf e}_{2}
\end{array} \right)_{x}= A_{p}
\left ( \begin{array}{cc}
{\bf e}_{1} \\
{\bf e}_{2}
\end{array} \right), \quad
\left ( \begin{array}{cc}
{\bf e}_{1} \\
{\bf e}_{2}
\end{array} \right)_{y}= B_{p}
\left ( \begin{array}{cc}
{\bf e}_{1} \\
{\bf e}_{2}
\end{array} \right), \quad
\left ( \begin{array}{cc}
{\bf e}_{1} \\
{\bf e}_{2}
\end{array} \right)_{t}= D_{p}
\left ( \begin{array}{cc}
{\bf e}_{1} \\
{\bf e}_{2}
\end{array} \right)
\eqno(104)
$$
where
$$
A_{p} =
\left ( \begin{array}{cc}
0             & k     \\
-\beta k      & 0
\end{array} \right) ,
\quad
B_{p}=
\left ( \begin{array}{cc}
0            & m_{3}   \\
-\beta m_{3} & 0
\end{array} \right)
$$
$$
D_{p}=
\left ( \begin{array}{cc}
0            & \omega_{3}   \\
-\beta \omega_{3} & 0
\end{array} \right).  \eqno(105)
$$

In the plane case the mM-LXII equation takes the following simple form
$$
k_{y} = m_{3x}   \eqno(106a)
$$
$$
k_{t} = \omega_{3x} \eqno(106b)
$$
$$
m_{3t}=\omega_{3y}.  \eqno(106c)
$$
Hence we get
$$
m_{3} = \partial^{-1}_{x}k_{y}. \eqno (107)
$$
The NEE has the form  (106b).
Many (2+1)-dimensional integrable equations such as the
Kadomtsev-Petviashvili, Novikov-Veselov (NV), mNV, KNV, (2+1)-KdV, mKdV
equations are the integrable reductions of
the M-LXII equation (106).
For example, let us show that the KP and  mKP equations are exact
reductions of the mM-LXII equation (106).
Consider the M-X equation [8]
$$
{\bf S}_t = \frac{\omega_{3}}{k}{\bf S}_{x}  \eqno(108)
$$
where
$$
\omega_{3}= -k_{xx}-3k^{2}--3\alpha^{2}\partial^{-1}_{x}m_{3y}. \eqno(109)
$$
If we put ${\bf S}= {\bf e}_{1}$ then  from (106) we obtain the L-equivalent
counterpart of the M-X equation which is the KP equation
$$
k_{t} +6kk_{x} + k_{xxx} +3\alpha^{2}m_{3y}=0 \eqno(110a)
$$
$$
m_{3x}=k_{y}. \eqno(110b)
$$
As known the LR of this equation is given by
$$
\alpha\psi_{y}+\psi_{xx}+k\psi = 0 \eqno(111a)
$$
$$
\psi_{t}+4\psi_{xxx}+6k\psi_{x}+3(k_{x}-\alpha m_{3})\psi =0. \eqno(111b)
$$
\subsubsection{The Zakharov and M-IX  equations}
Now we find the connection between the Myrzakulov IX (M-IX) equation
and the curves (the M-LXI equation). The M-IX equation reads as
$$
{\bf S}_t = {\bf S} \wedge M_1{\bf S}+A_2{\bf S}_x+A_1{\bf S}_y  \eqno(112a)
$$
$$
M_2u=2\alpha^{2} {\bf S}({\bf S}_x \wedge {\bf S}_y) \eqno(112b)
$$
where $ \alpha,b,a  $=  consts and
$$
M_1= \alpha ^2\frac{\partial ^2}{\partial y^2}+4\alpha (b-a)\frac{\partial^2}
   {\partial x \partial y}+4(a^2-2ab-b)\frac{\partial^2}{\partial x^2},
$$
$$
M_2=\alpha^2\frac{\partial^2}{\partial y^2} -2\alpha(2a+1)\frac{\partial^2}
   {\partial x \partial y}+4a(a+1)\frac{\partial^2}{\partial x^2},
$$
$$
A_1=i\{\alpha (2b+1)u_y - 2(2ab+a+b)u_{x}\},
$$
$$
A_2=i\{4\alpha^{-1}(2a^2b+a^2+2ab+b)u_x - 2(2ab+a+b)u_{y}\}. \eqno (113)
$$
The M-IX equation was introduced in [8] and is integrable. It admits several
integrable reductions: 1) the Ishimori equation as $a=b=-\frac{1}{2}$;
2) the M-VIII equation as $a=b=-1$ and so on [8]. In this case we have
$$
m_{1}=\partial_{x}^{-1}[\tau_{y}-\frac{\beta}{2\alpha^2}M_2 u] \eqno(114a)
$$
$$
m_{2}=-\frac{1}{2\alpha^2 k}M_2 u  \eqno (114b)
$$
$$
m_{3}=\partial_{x}^{-1}[k_y +\frac{\tau}{2\alpha^2 k}M_2 u]  \eqno(114c)
$$
and
$$
\omega_{1} = \frac{1}{k}[-\omega_{2x}+\tau\omega_{3}],
                 \eqno (115a)
$$
$$
\omega_{2}= -4(a^{2}-2ab-b)k_{x}-
4\alpha (b-a)k_{y} -\alpha^{2}(m_{3y}+m_{2}m_{1})+m_{2}A_{1}
\eqno (115b)
$$
$$
\omega_{3}= -4(a^{2}-2ab-b)k \tau-
4\alpha (b-a)k m_{1}+\alpha^{2}(m_{2y}-m_{3}m_{1})
+k A_{2}+m_{3}A_{1}.
\eqno (115c)
$$
Functions $q, p$ are given by (96) with
$$
a_{1}^2 =\frac{|a|^2}{|b|^2}a_{1}^{\prime^{2}}=\frac{|a|^2}{|b|^2}\{(l+1)^2k^2
+\frac{|\alpha|^2}{4}(m_3^2 +m_2^2)-(l+1)\alpha_{R}km_3-
(l+1)\alpha_{I}km_2\}
  \eqno(116a)
$$
$$
b_{1} =\partial_{x}^{-1}\{-\frac{\gamma_1}{2ia_1^{\prime^{2}}}-(\bar A-A+D-\bar D)\}
\eqno(116b)
$$
$$
a_{2}^2 =\frac{|b|^2}{|a|^2}a_{2}^{\prime^{2}}=\frac{|b|^2}{|a|^2}\{l^2k^2
+\frac{|\alpha|^2}{4}(m_3^2 +m_2^2)-l\alpha_{R}km_3+
l\alpha_{I}km_2\}
  \eqno(116c)
$$
$$
b_{2} =\partial_{x}^{-1}\{-\frac{\gamma_2}{2ia_2^{\prime^{2}}}-(A-\bar A+\bar D-D)
\eqno(116d)
$$
where
$$
\gamma_1=i\{2(l+1)^2k^{2}\tau+\frac{|\alpha|^2}{2}(m_3km_1+m_2k_y)-
$$
$$
(l+1)\alpha_{R}[k^{2}m_1+m_3k\tau+
m_2k_x]+(l+1)\alpha_{I}[k(2k_y-m_{3x})-
k_x m_3]\} \eqno(117a)
$$
$$
\gamma_2=-i\{2l^2k^{2}\tau+
\frac{|\alpha|^2}{2}(m_3km_1+m_2k_y)-
$$
$$
l\alpha_{R}(k^{2}m_1+m_3k\tau+
m_2k_x)-l\alpha_{I}[k(2k_y-m_{3x})-
k_x m_3]\}. \eqno(117b)
$$
Here $\alpha=\alpha_{R}+i\alpha_{I}$. In this case, $q,p$
satisfy the following other Zakharov equation [49]
$$
iq_t+M_{1}q+vq=0 \eqno(118a)
$$
$$
ip_t-M_{1}p-vp=0 \eqno(118b)
$$
$$
M_{2}v=-2M_{1}(pq) \eqno(118c)
$$
which is integrable and admits several particular cases.

As well known the M-IX equation admits several reductions:
1)  the M-IXA equation as $\alpha_{R}=1, \alpha_{I}=0$;
2)  the M-IXB equation as $\alpha_{R}=0, \alpha_{I}=1$;
3) the M-VIII equation as $a=b=1$
4) the IE $a=b=-\frac{1}{2}$
and so on. The corresponding versions of the ZE (118), we obtain as
the corresponding values of the parameter $\alpha$.

\section{Conclusion}

In this paper we analyzed the M-LXVIII equation. We have found
the some integrable reductions of this equation. Also we have shown  that
the Zakharov and its spin counterpart the M-I equation are exact reductions
of the SDYME. The higher hierarchy of the SDYME was introduced. Using
this hierarchy it was shown that several simplest soliton equations
in 2+1 dimensions such as mKdV, derivative NLS and M-III$_{q}$ equations
and so on are also its exact reductions.

Finally we would like ask you, dear colleaque, if you know, have or will
have any results on multidimensional soliton equations, soliton geometry
and the Yang-Mills equations, please, inform me (R.M.) or send me a hard copy
of your papers. Also any comments, proposals and questions are welcome.

\section{Acknowledgments}

R.M. is grateful to M.Lakshmanan for hospitality during visits,
many useful discussions and for financial support. Special thanks to Radha
Balakrishnan and M.Daniel for discussions. We thank G.N.Nugmanova for
helpful comments on the manuscript.

\section{Tasks}
Task-1: Please find the breaking solutions (instantons, monopols,
dions and so on ) of the SDYME.
\\
Task-2: Please find the breaking solutions of the Bogomolny equation.
\\
Task-3: Please find the breaking solutions of the Ishimori and DS equations.
\\
Task-4: Please find the breaking solutions of the M-IX  and Zakharov equations.
\\
Task-5: Please consider the above presented results from twistor point of view.

\end{document}